\documentclass[twocolumn,english,prl]{revtex4}
\usepackage[latin9]{inputenc}
\usepackage{amssymb}
\usepackage{graphicx}
\usepackage{amsmath}
\usepackage{color}
\usepackage{mathrsfs}
\usepackage{float}
\usepackage{indentfirst}
\usepackage{mathrsfs}
\usepackage{float}
\usepackage{indentfirst}
\usepackage{textcomp}
\usepackage{comment}
\usepackage{mathtools}

\newcommand{\COMMENTED}[1]{}

\begin{document}

\title{Rashba spin-orbit coupling, strong interactions, and the BCS-BEC crossover \\ in the ground state of the two-dimensional Fermi Gas}

\author{Hao Shi}
\author{Peter Rosenberg}
\author{Simone Chiesa}
\author{Shiwei Zhang}

\affiliation{Department of Physics,
             The College of William and Mary,
             Williamsburg, Virginia 23187}

\begin{abstract}
The recent experimental realization of spin-orbit coupled Fermi gases provides a unique opportunity to study the 
interplay between strong interaction and SOC in a tunable, disorder-free system. 
We present here precision ab initio numerical results on the two-dimensional, unpolarized, uniform Fermi gas with attractive interactions and Rashba SOC. Using auxiliary-field quantum Monte Carlo and incorporating recent algorithmic advances, we carry out exact calculations on sufficiently large system sizes to provide accurate results systematically as a function of experimental parameters. We obtain the equation of state,
the momentum distributions, the pseudo-spin correlations and the pairing wave functions. 
Our results help illuminate the rich pairing structure induced by SOC, and provide benchmarks for theory and guidance to future experimental efforts.
\end{abstract}


\maketitle

Spin-orbit coupling (SOC) plays a fundamental role in a number of physical contexts spanning nuclear, atomic, and condensed 
matter physics. SOC in two-dimensional (2D) systems is particularly relevant to condensed matter physics, because of connections to
the quantum Hall effect, and topological insulators and superconductors, among others. While it can be difficult to isolate and study the effects of SOC in typical condensed matter settings, the advent of synthetic gauge fields in ultracold atomic gases\,\cite{SOC_BEC_2009,SOC_BEC_2011,SOC_DFG_CHEUK,SOC_DFG_WANG} provides unprecedented access to clean, tunable systems in which it is possible to precisely
investigate the interplay between interaction and SOC. Current experimental efforts have primarily achieved a combination of Rashba and Dresselhaus SOC. Recently, pure Rashba SOC was realized using a three laser Raman scheme\,\cite{SOC_RASHBA_2015}, 
and a number of proposals exist for dark-state, generalized Raman, and magnetic schemes 
\cite{PhysRevLett.99.110403, PhysRevA.81.053403, PhysRevA.77.011802, PhysRevA.84.025602, PhysRevB.83.140510, PhysRevLett.111.125301, 
PhysRevA.87.063634, 2015arXiv151101588C}.
 
Even without SOC, Fermi gas systems have been a fertile ground for fundamental advances in many-body physics. 
The precise agreement achieved between experiment and theory in three-dimensions is a triumph for understanding strongly correlated Fermion systems
\cite{Ku563,Kinast1296,LeLuoJLTP,PhysRevA.84.061602,PhysRevA.79.013627}.
Recently,
the 2D Fermi gas has drawn considerable attention
\cite{2DFG_AFQMC,2DFG_EXPT,2DFG_DMC,PhysRevA.77.053617,PhysRevA.78.043617,PhysRevA.67.031601,PhysRevA.77.063613,PhysRevLett.95.170407,PhysRevA.67.031601}, 
for the possibility to study with great precision fermion pairing in 2D, which is important 
in high-$T_c$ and other exotic matter. 
SOC adds a new 
layer of complexity to the rich pairing picture,
with the presence of both singlet and triplet pairing, and
the interplay with spin chirality.

These recent experimental advances have thus prompted intense 
theoretical efforts to study SOC in the 2D Fermi gas, many of which focus on the 
connection between SOC and the BCS-BEC crossover \cite{PhysRevA.85.013601,PhysRevLett.107.195305,PhysRevLett.107.195304,PhysRevLett.108.145302,MFT_Shenoy,PhysRevA.85.011606}. However, as is commonly the case in the study of strongly interacting systems, mean-field theory is often the only available tool.
To date almost all the theoretical and computational work on the Fermi gas 
has been done at the mean-field level. 
It is therefore crucial to understand and quantify the corrections from particle correlations, in order to validate the predictions from mean-field calculations.
Establishing precise benchmark results is also of fundamental value in guiding and calibrating 
experiments and assessing new theoretical and computational methods 
as they are developed for treating SOC in the presence of strong interactions.

In this work we present the first exact results on the ground state of the 2D Fermi gas with strong attractive interactions and 
Rashba SOC. We show how SOC effects in many-fermion systems can be treated by auxiliary-field quantum Monte Carlo (AFQMC),
formulated as random walks of general Slater determinants consisting of spin-orbitals.
The method can be generalized to carry out \emph{ab initio} calculations in real materials which will be important in the 
investigation of novel phases of matter under the interplay of topological physics and strong electron correlations. 

For the unpolarized 2D Fermi gas with SOC, this method allows numerically exact calculations free of the sign problem. 
Combining it with Monte Carlo algorithmic advances, we are able to simulate large lattice sizes to reach the ground state and 
the continuum limit, and sufficiently large number of particles to reach the thermodynamic limit. An equation of state is obtained 
which can serve as a benchmark for future theory and experimental efforts. The correlation energy is found to be nearly independent 
of SOC strength. We also present a detailed study of the momentum distributions, pseudo-spin correlations, the singlet and triplet pairing wave functions, and the condensate fractions as a function of SOC and interaction strengths. The results present a precision benchmark for an exotic quantum system which, on the verge of experimental realization, combines topological effects and superconductivity.

The Hamiltonian for the 2D Fermi gas with attractive zero-range interactions and Rashba SOC can be written as a sum of three pieces,
\begin{equation}
\hat{H}=\hat{H}_0+\hat{H}_\textrm{SOC}+\hat{H}_\textrm{int},
\label{eq:Hamiltonian}
\end{equation}
which correspond to the kinetic, SOC, and interaction energy.
We consider $N$ particles in a periodic box, represented on a lattice of dimension $L\times L$, so that
\begin{align}
&\hat{H}_0=\sum_{\mathbf{k},\sigma} \varepsilon_\mathbf{k} c^{\dagger}_{\mathbf{k}\sigma} c_{\mathbf{k}\sigma},\notag\\
&\hat{H}_\textrm{SOC}= \sum_{\mathbf{k}}\lambda \left(k_y-i k_x\right) c^{\dagger}_{\mathbf{k}\downarrow} c_{\mathbf{k}\uparrow}+ h.c.,\notag\\
&\hat{H}_\textrm{int}=U\sum_{\mathbf{i}} n_{\mathbf{i}\uparrow} n_{\mathbf{i}\downarrow},   
\label{eq:Hamiltonian_terms}
\end{align}
where $c^{\dagger}_{\mathbf{k}\sigma}$ is the creation operator for a fermion with spin $\sigma$ and momentum $\mathbf{k}$.
The number operators on lattice site $\mathbf{i}$ are $n_{\bold{i}\sigma}=c^\dagger_{\bold{i}\sigma} c_{\bold{i}\sigma}$, and
the dispersion relation is $\varepsilon_\mathbf{k}=\vert \mathbf{k}\vert^2=(k_x^2+k_y^2)$. 
The Hamiltonian in Eq.~(\ref{eq:Hamiltonian}) can be directly mapped to the continuum form (e.g., as in experiments) by an overall energy scale defined by the ground-state energy per particle of the corresponding non-interacting Fermi gas,  $E_{FG}$ (which in the present form is $\pi n$, with $n=N/L^2$ the number density). The interaction strength $U$ is uniquely defined \cite{WernerCastin,2DFG_AFQMC} by $\log(k_F a)$ where the Fermi wave-vector $k_F$ measures the inverse of the average inter-particle spacing while $a$ is the scattering length. 
It is convenient to introduce two dimensionless parameters:
\begin{align}
\alpha = \frac{\lambda^2}{E_{FG}}; \quad
\beta = \frac{\varepsilon_B}{E_{FG}},
\end{align}
to specify the strengths of the SOC and interaction, respectively,
where $\varepsilon_B$ is the two-body binding energy at $\lambda=0$ and is directly related to $k_F a$ \cite{2DFG_AFQMC}.

Our calculations treat  periodic lattices of over 1200 sites, typically with over 70 fermions.  
For each set of parameters, the many-body  ground state is computed  
using the AFQMC framework \cite{Lecture-notes,BSS,Koonin}, generalized to treat SOC.
In AFQMC, one  projects out the ground state of $\hat H$ from an initial state $|\phi^{(0)}\rangle$
by repeated applications of the imaginary-time propagator $e^{-\tau \hat H}$, which
is decoupled into path integrals over independent-particle propagators defined by auxiliary-fields. 
The path integrals can be evaluated by Monte Carlo, which can be realized as
random walks in the space of Slater determinants, starting from $|\phi^{(0)}\rangle$.
Without SOC, each Slater determinant takes the form of a Hartree-Fock solution,
  $|\phi\rangle=|\phi_\uparrow\rangle \otimes |\phi_\downarrow\rangle$, 
where the $\uparrow$- and $\downarrow$-spin components are  $N_s \times N_\uparrow$ and $N_s \times N_\downarrow$ matrices, respectively,
with $N_s$ being the basis size ($=L^2$ here) and $N_\sigma$ being the number of $\sigma$-spin fermions ($=N/2$ here). 
With SOC, this must be replaced with the generalized  Hartree-Fock form, of a $2N_s\times N$ matrix. The matrix elements evolve stochastically,
being propagated by one-body propagators which sample auxiliary-fields and each of which can be thought as an $2N_s\times 2N_s$ matrix.

The Fermi gas Hamiltonian, with $\lambda=0$, is free of the sign problem, because $|\phi_\uparrow\rangle$ can be made identical to $|\phi_\downarrow\rangle$ for every random walker, so that the trace or ground-state overlap over each path has the form of the square of a determinant 
and is thus non-negative. With SOC, it is straightforward to show that time-reversal symmetry is preserved, ${\hat T} {\hat H_{\rm SOC}} {\hat T^{-1}}={\hat H_{\rm SOC}}$, as is already the case with $\hat H_0$ and $\hat H_{\rm int}$. Thus there is no sign problem \cite{nosignproblem,determinant_QMC_oplatt}, with the eigenvalues of the overlap matrix being 
complex-conjugate pairs and thereby the determinant being non-negative ~\cite{note-odd-lattice-size-plusk-pts}. 
(Of course the $\lambda=0$ Hamiltonian can be viewed as a special case, by thinking of $|\phi_\uparrow\rangle$ and $|\phi_\downarrow\rangle$
as two diagonal blocks of the $2N_s\times N$  supermatrix.) We apply  dynamic force biases \cite{2DFG_AFQMC} in sampling the AF paths
to achieve high efficiency. All numerical biases or systematic errors in the calculations have been controlled so that they are smaller than our statistical uncertainty. The high-precision results obtained are therefore fully {\it ab initio\/} and are exact for each parameter set.

\begin{figure}
\includegraphics[width=\columnwidth]{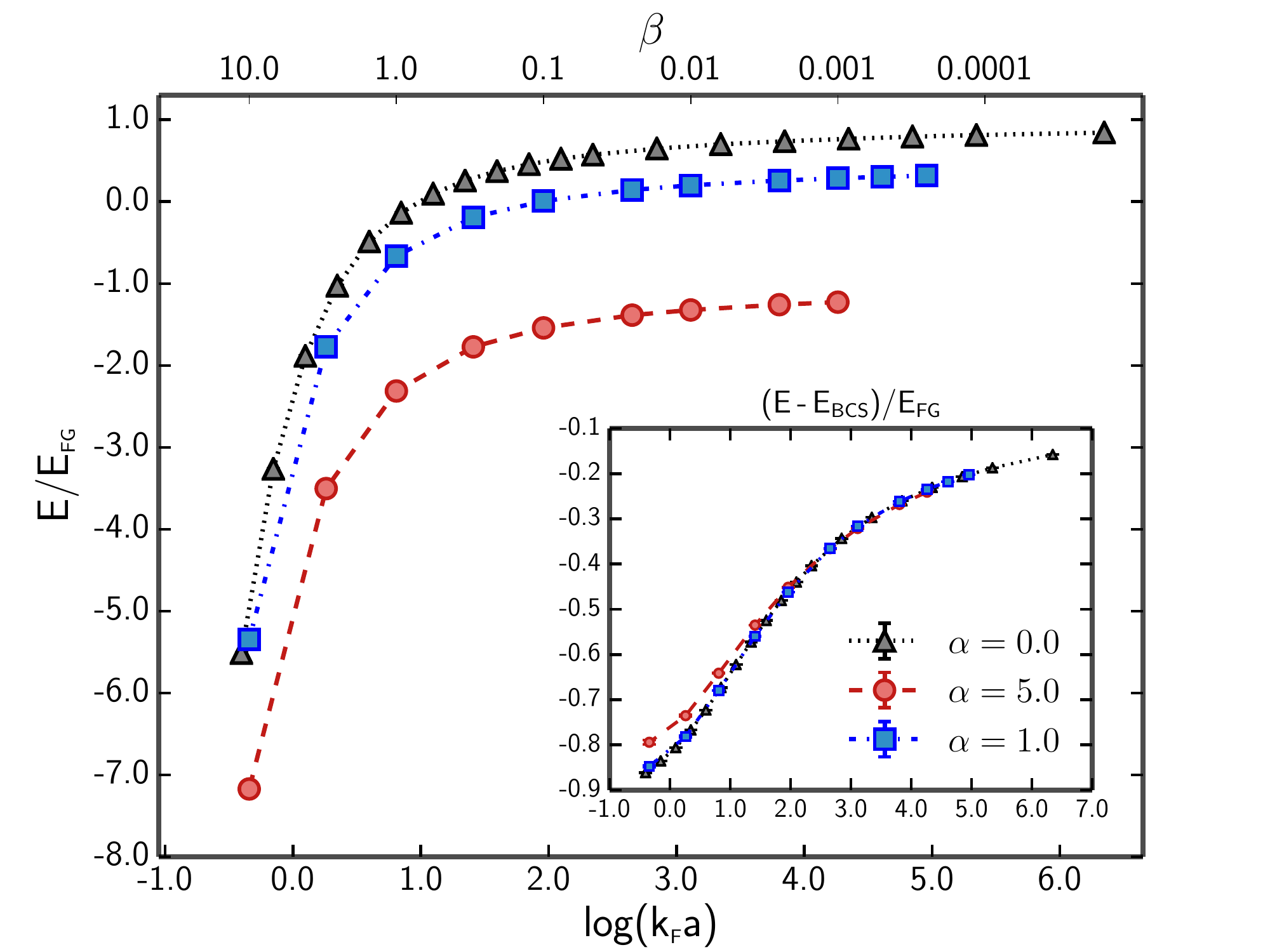}
\caption{Equation of state for three SOC strengths, $\alpha=0.0$ (triangle), 1.0 (square), and 5.0 (circle). Results have been extrapolated to
the continuum and  thermodynamic limit. The inset plots the results relative to those from
BCS, revealing that the correlation energy is quite insensitive to SOC strength.}
\label{fig:EOS}
\end{figure}

In Figure \ref{fig:EOS} we present the computed equation of state as a function of interaction strength, $\log(k_Fa)$, for several values of SOC strength. The results are first extrapolated to the continuum limit with calculations on a sequence of $L$ values with $N$ fixed, and then larger $N$ systems are computed until convergence is obtained \cite{2DFG_AFQMC}. 
Results for the 2D FG without SOC \cite{2DFG_AFQMC} are also shown as a reference. 
The most dramatic effect of SOC is a decrease of the total energy, which plateaus at large $\log(k_Fa)$.
This shift to the energy becomes more pronounced at larger values of SOC strength. The inset of Fig.\,\ref{fig:EOS} displays the difference between the QMC energy and the energy predicted by BCS theory. This difference 
provides a measure of the correlation energy. The similarity in the behavior of the curves suggests that the correlation energy is relatively insensitive to SOC, with a small effect becoming noticeable for systems
with strong SOC, in the crossover or BEC regime.

The non-interacting part of the Hamiltonian can be expressed in diagonal form in the helicity basis 
with the corresponding dispersion relations, $\varepsilon^\pm_\mathbf{k}=\mathbf{k}^2\pm\lambda|\mathbf{k}|$.
We examine the properties of the many-body ground state in this representation by working in natural orbital space.
We diagonalize the one-body density matrix,
\begin{equation}
\begin{pmatrix}
\langle n_{\mathbf{k}\uparrow}\rangle && \langle S^+_{\mathbf{k}}\rangle \\
\langle S^-_{\mathbf{k}}\rangle && \langle n_{\mathbf{k}\downarrow}\rangle 
\end{pmatrix}
=
\begin{pmatrix}
\langle c^\dagger_{\mathbf{k}\uparrow} c_{\mathbf{k}\uparrow}\rangle && \langle c^\dagger_{\mathbf{k}\uparrow} c_{\mathbf{k}\downarrow}\rangle \\
\langle c^\dagger_{\mathbf{k}\downarrow} c_{\mathbf{k}\uparrow}\rangle && \langle c^\dagger_{\mathbf{k}\downarrow} c_{\mathbf{k}\downarrow}\rangle 
\end{pmatrix},
\end{equation}
where the expectation values are taken with respect to the many-body ground state.
The eigenvalues yield the momentum distribution in the helicity bands, $n^\pm_\mathbf{k}$. 
The spin orientation is specified by $(S^x, S^y)$, which are computed from $\langle S^\pm_{\mathbf{k}}\rangle$ directly.

\begin{figure}
\includegraphics[width=\columnwidth]{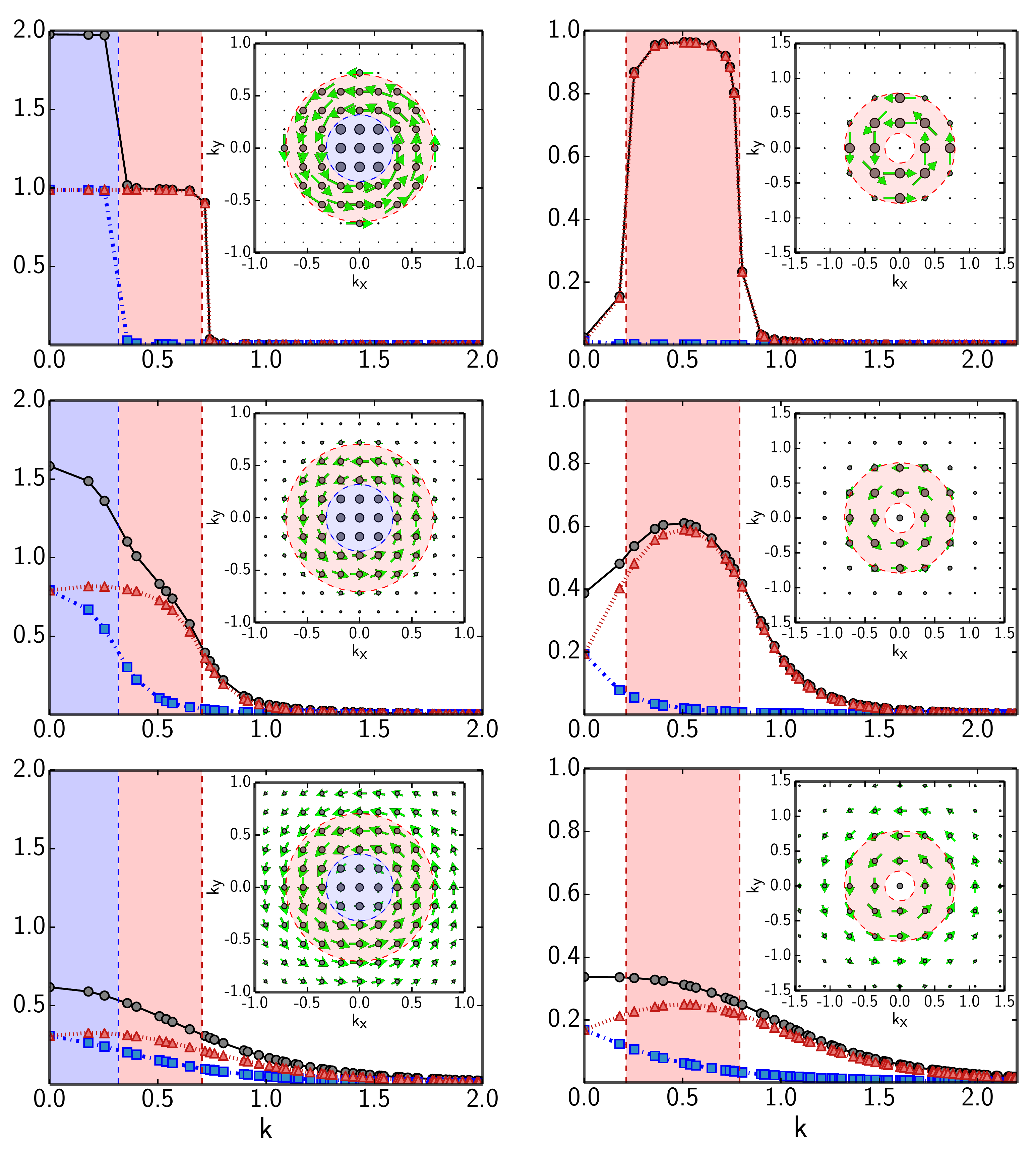}
\caption{Momentum distributions, $n^+_\mathbf{k}$ (squares), $n^-_\mathbf{k}$ (triangles), $n^\textmd{total}_\mathbf{k}$ (circles) for 
modest ($\alpha=1.0$,
left column) and strong ($\alpha=7.0$,
right column) SOC.  
 From top to bottom, the rows correspond to weak ($\beta=0.001$), intermediate ($\beta=1.0$), and strong ($\beta=10.0$) interaction strength. The non-interacting Fermi surfaces are indicated by the 
 vertical dashed lines, and the occupation for each band is indicated by the corresponding shaded regions (in both the main plot and the inset). In the insets, the arrows point to the direction of $\langle \mathbf{S}_\mathbf{k}\rangle$, and their size indicate its magnitude.
The size of the dots represents the magnitude of $n^\textmd{total}_\mathbf{k}$. 
These calculations had $L=35$ and $N=58$ (left column) and  $N=56$ (right column).
(Note that different scales are used between the two columns, and between the 
last row and the other two to improve clarity.)
}
\label{momentum_plus_spin}
\end{figure}

\begin{figure*}
  \includegraphics[width=\textwidth]{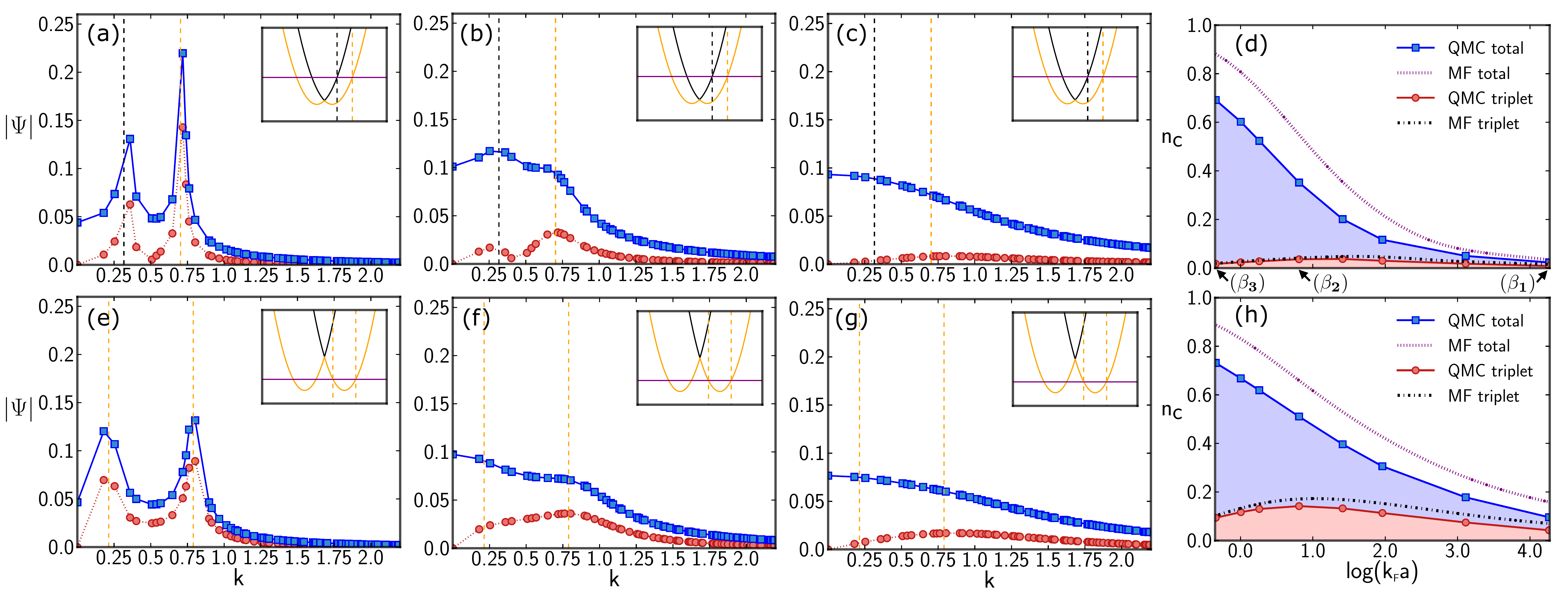}
   \caption{Singlet (square) and triplet (circle) components of the condensate wave function, and the condensate fraction.
(a)-(d) are for $\alpha = 1.0$ and  (e)-(h) are for $\alpha = 7.0$.
 The first three panels in each row show the wave functions at increasing interaction strength ($\beta_1=0.001$, $\beta_2=1.0$, and $\beta_3=10.0$,
 values indicated by arrows in panel (d)).
The insets show the helicity bands, $\varepsilon^\pm_\mathbf{k}$, 
   and the non-interacting Fermi surfaces, indicated by the vertical dashed lines. 
   The systems are the same as in Fig.~\ref{momentum_plus_spin}.
   }
   \label{pair_plus_cond_frac}
\end{figure*}

Plotted in Fig.~\ref{momentum_plus_spin} are the momentum distributions
for each helicity band at several values of interaction strength.
The insets show the pseudo-spin orientation and magnitude.
The helicity bands and the non-interacting Fermi surfaces are indicated for reference. 
(They are also 
illustrated in more detail in the insets in Fig.~\ref{pair_plus_cond_frac}.)
In the weak SOC regime, both helicity bands are occupied, while for strong SOC only the $\varepsilon^-_\mathbf{k}$ band is occupied.
The transition between the two
is at $\alpha =4.0$ for $\beta=0$. Our calculations indicate a smooth transition in the presence of interaction.

At small interaction strengths the momentum distributions 
deviate very little from the non-interacting case, as expected.
As $\beta$ increases, the sharper features of the momentum distributions smoothen and the distributions broaden, 
indicating that higher momentum states have become occupied. At intermediate and large interaction strengths the discrepancy
from the non-interacting case becomes quite apparent, as interaction dramatically alters the structure defined by the shaded regions.
For large SOC, for instance, both bands become occupied and lower $\mathbf{k}$ states, which are empty in the non-interacting case,
 are heavily populated.

We next examine the pairing properties of the system as a function of 
SOC and interaction strength. We focus on the interplay of singlet and triplet pairing, and connect the pairing structure to the condensate wave function and condensate fraction. With the pairing operators
\begin{align}
\Delta^\dagger_\uparrow(\mathbf{k})&=c^\dagger_{\mathbf{k}\uparrow}c^\dagger_{\mathbf{-k}\uparrow};
\quad 
\Delta^\dagger_\downarrow(\mathbf{k})=c^\dagger_{\mathbf{k}\downarrow}c^\dagger_{\mathbf{-k}\downarrow};\notag\\
\Delta^\dagger_s(\mathbf{k})&=\frac{1}{\sqrt{2}}\left(c^\dagger_{\mathbf{k}\uparrow}c^\dagger_{\mathbf{-k}\downarrow}-
c^\dagger_{\mathbf{k}\downarrow}c^\dagger_{\mathbf{-k}\uparrow}\right)\,,
\label{eqn:pair_creation_ops}
\end{align}
we construct the following $3N_s \times 3N_s$ zero-momentum pairing matrix
\begin{equation}
M_{\sigma\sigma^\prime}(\mathbf{k},\mathbf{k}^\prime)=
\langle\Delta^\dagger_\sigma(\mathbf{k})\Delta_{\sigma^\prime}(\mathbf{k}^\prime)\rangle,
\label{eq:pm_3Lx3L}
\end{equation}
with $\sigma,\sigma^\prime=\,\uparrow,\downarrow$, or $s$.
The leading eigenvalue, $N_c$, of the pairing matrix yields the condensate fraction, $n_c\equiv N_c/N$.
The corresponding eigenstate gives the condensate wave function
in $\mathbf{k}$-space \cite{Yang1962}. 
The condensate wave function is composed of singlet and triplet components,
$|\Psi_c\rangle = |{\Psi_c}_{, s}\rangle + |{\Psi_c}_{, t}\rangle$.
With $|\Psi_c\rangle$ normalized,
we define the singlet and triplet contributions to the condensate fraction by
$n_{c,s}/n_c=\langle {\Psi_c}_{,s} | {\Psi_c}_{,s} \rangle$ and $n_{c,t}/n_c=\langle \Psi_{c,t} | \Psi_{c,t} \rangle$ respectively.

The singlet and triplet components of the condensate wave function, and the condensate fraction, are plotted for
several representative values of SOC and interaction strength in Fig.~\ref{pair_plus_cond_frac}. The anti-symmetry of
the triplet wave function is reflected by the presence of a node at $\mathbf{k}=0$, while the 
symmetric singlet component has no node. 

As SOC strength increases, the amplitude of the triplet component of the wave function becomes closer to that  
of the singlet, and the triplet portion of the condensate fraction grows relative to the singlet component.  
The total condensate fraction grows with SOC strength, primarily a consequence of the increase in 
triplet pairing, which is induced by SOC and vanishes as $\alpha\rightarrow0$. BCS theory tends to over-estimate both 
components but is seen to especially over-estimate the singlet component.

As interaction strength increases the sharp 
peaks of the wave function, which occur in the vicinity of the Fermi surface, broaden and become smooth. 
While pairing is confined to the Fermi surface at weak interactions, the large modifications to the momentum distributions at strong interactions
cause pairing to occur over a wide range of momenta, including a peak in the singlet component at low $|\mathbf{k}|$,
centered around states which are un-occupied in the independent-particle picture.
The pairing wave functions in (a) exhibit larger peaks on the right (at larger $|\mathbf{k}|$), in contrast with 
two relatively even peaks in (e). This is a consequence of the very different behaviors of the 
momentum distribution. 
For $\alpha=7.0$, 
many unoccupied momentum states are available in the vicinity of the
Fermi surface at lower $|\mathbf{k}|$ 
to facilitate pairing, which is not the case for $\alpha=1.0$.

The shape and amplitude of the singlet and triplet components of the condensate wave function are most similar at small interaction 
strength, and the contributions to the condensate from singlet and triplet pairs are of roughly equal magnitude. For large interaction strength, 
the amplitude of the triplet wave function is significantly reduced and the condensate fraction is primarily composed of singlet pairs.  
The triplet component of the condensate fraction has a peak around $\log(k_Fa) = 1.0$ suggesting that triplet pairing is maximized in the crossover regime, where the strength of the interaction is large enough to induce robust pairing, but not so large as to discourage triplet pair formation.

\begin{figure}
\includegraphics[width=\columnwidth]{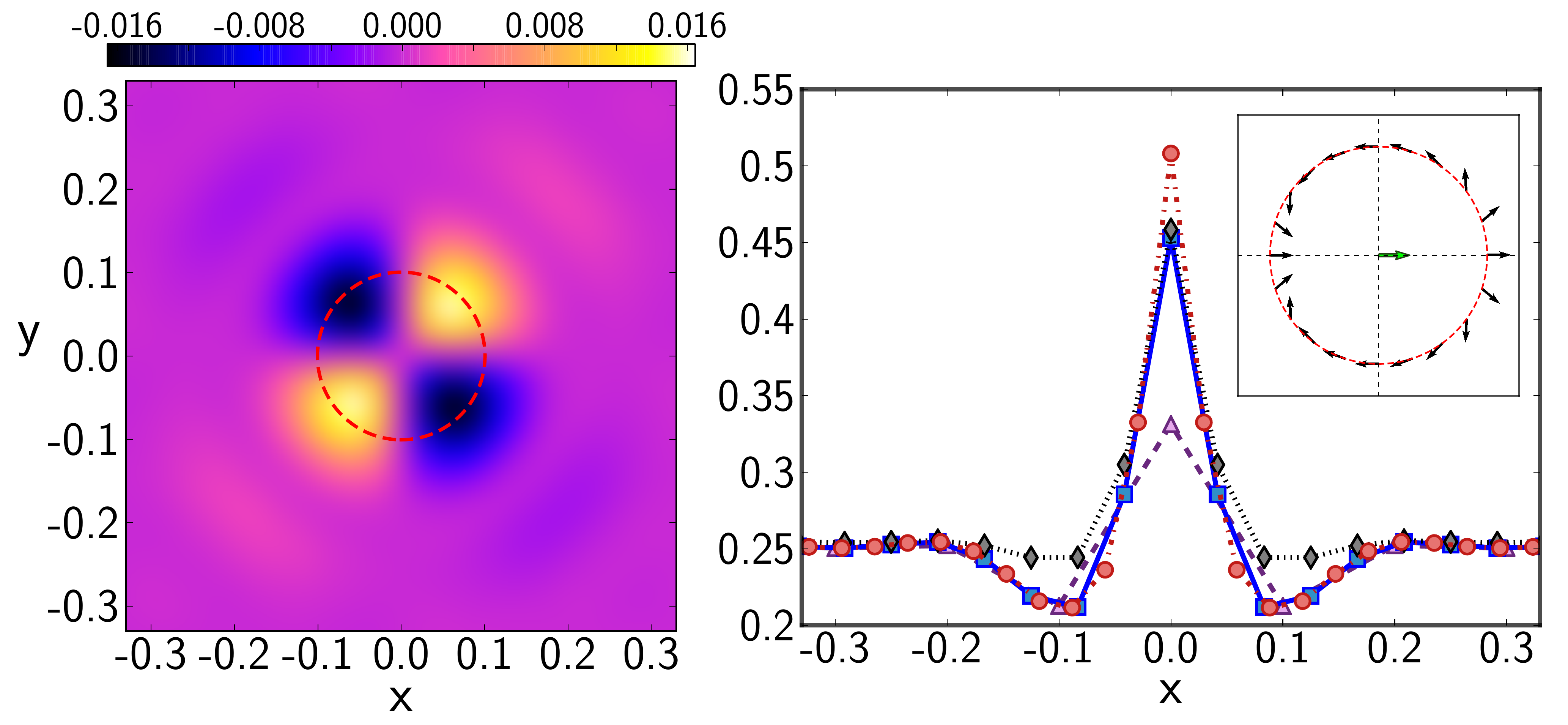}
\caption{Real-space pairing structure, nematic order, and spin 
chirality index.
 Plotted on the left is $\langle Q^{xy}(0,r)\rangle$ 
 for $\alpha=1.0$, $\beta=0.001$, with average inter-particle spacing $1/k_F = 0.0524$. 
The right panel shows the (isotropic) spin correlation $\langle n_{0\uparrow}n_{r\downarrow}\rangle$ for $L=$11 (purple triangle), 
25 (blue square), and 35 (red circle). 
The black diamonds plot a reference curve without SOC 
for $L=25$.
The inset illustrates the
chirality
of the pair along the dashed red circle shown in the plot of $Q^{xy}$.
}
\label{fig:Qxy}
\end{figure}

To probe the  real-space structure of pairs and examine possible spin nematic order  in the presence of Rashba SOC, we compute the spin correlator
defined as \cite{spin_nematic_order},
\begin{equation}
\hat{Q}^{ij}(\mathbf{r}_1,\mathbf{r}_2) = \frac{1}{2}\left(\hat{S}^i_1\hat{S}^j_2+\hat{S}^j_1\hat{S}^i_2\right)-\frac{\delta^{ij}}{3}\hat{\mathbf{S}}_1\cdot\hat{\mathbf{S}}_2,
\label{eq:quad_operator}
\end{equation}
where
the subscript refers to $\mathbf{r}_1$ or $\mathbf{r}_2$ and $i$ and $j$ denote $x$, $y$, and $z$. 
As depicted in Fig.~\ref{fig:Qxy},
$\langle Q^{xy} \rangle$ (and similarly, $\langle Q^{xx} \rangle$) yields a flower-shaped pattern, 
a $4\pi$ rotation of the second spin in the pair, relative to the first spin, along a circular path around the origin. This spin rotation is illustrated in the upper right panel of Fig.~\ref{fig:Qxy}, which gives the direction of the spin along the dashed red circle in the plot of $\langle Q^{xy} \rangle$.  
Similar chirality/winding behaviors have been observed in pseudo-spins in layered materials 
\cite{PhysRevLett.115.176801,ncomms4876,PhysRevLett.96.086805,PhysRevB.77.041407}. 
SOC causes a dramatic difference in the spin correlation as shown in the right panel.
With SOC turned on, a significant 
decrease in  $\langle n_{0\uparrow}n_{r\downarrow}\rangle$ is seen immediately beyond the central peak.
However the total density-density correlation (not shown) is essentially unchanged.
This signals a decrease in singlet pairing which is compensated for by an increase in triplet pairing. 

In summary, we have developed an approach for exact numerical computations of  the ground state of the strongly interacting 
Fermi gas under SOC, and have provided the first systematic results beyond mean-field theory. A detailed equation of state is obtained.
The correlation energy is seen to be nearly independent of SOC strength. Dramatic deviations are seen from the non-interacting picture in the momentum distribution. The condensate fraction is computed. Triplet pairing appears under SOC, and the interplay between interaction and SOC causes triplet pairing to be maximized in the crossover region. Nematic correlation develops but no long-range order is seen. A spin chirality 
of 4$\pi$ is seen in the pair state. These {\it  ab initio\/} precision many-body results provide benchmark for theory and can serve as a calibration for experiments.

\begin{acknowledgments}

We thank L.~He, E.~Rossi, P. Xu, C. Zhang, R. Zhang for useful discussions.
This research was supported by NSF
(grant no.~DMR-1409510), and the Simons Foundation. Computing was carried out at the
Oak Ridge Leadership Computing Facility at the Oak Ridge National Laboratory,
and at the computational facilities at the College of William and Mary.

\end{acknowledgments}

\bibliography{AUQMCBIB}

\begin{thebibliography}{45}
\bibitem[{not()}]{note-odd-lattice-size-plusk-pts}
\bibinfo{note}{Our calculations are performed on systems with a closed-shell
  and lattices with odd $L$ to preserve k-point symmetry.}
\end{thebibliography}

\section{Supplementary Material:}

\subsection{Equation of state data}
We list in Table~\ref{table:EOS_data} the equation of state data for $\alpha=1.0, 5.0$.
The reference data for $\alpha=0.0$ can be found in \cite{2DFG_AFQMC}.
\begin{table}[ht!]
\begin{tabular}{ccllcccccc}
\hline\hline
\multicolumn{1}{c}{$\log(k_Fa)$} && \multicolumn{1}{c}{$\beta$} && \multicolumn{1}{c}{$\alpha$} && \multicolumn{1}{c}{$E_\textmd{BCS}/E_\textmd{FG}$} && \multicolumn{1}{c}{$E_\textmd{QMC}/E_\textmd{FG}$} \\ \hline
4.956100  && 0.00025 && 	1.0  && 	0.521454  && 	0.318(2)  \\
4.609530  && 0.0005 && 	1.0  && 	0.521245  && 	0.303(2)   \\
4.262960  && 0.001 && 	1.0  && 	0.520805  && 	0.285(2)   \\
3.804810  && 0.0025 && 	1.0  && 	0.519664  && 	0.258(4)   \\
3.111660  && 0.01 && 	1.0  && 	0.515351  && 	0.199(5)   \\
2.653520  && 0.025 && 	1.0  && 	0.507885  && 	0.143(7)   \\
1.960370  && 0.1 && 	1.0  && 	0.470332  && 	0.008(5)   \\
1.411070  && 0.3 && 	1.0  && 	0.369147  && 	-0.190(5)  \\
0.809079  && 1.0 && 	1.0  && 	0.016098  && 	-0.663(4)  \\
0.259773  && 3.0 && 	1.0  && 	-0.988668  && 	-1.770(4)  \\
-0.342214  && 10.0 && 	1.0  && 	-4.494136  && 	-5.346(4)  \\
\\[-2ex]
4.262960  && 0.001 && 	5.0  && 	-0.985105  && 	-1.227(2)   \\
3.804810  && 0.0025 && 	5.0  && 	-0.988789  && 	-1.258(4)   \\
3.111660  && 0.01 && 	5.0  && 	-1.001324  && 	-1.323(5)   \\
2.653520  && 0.025 && 	5.0  && 	-1.019979  && 	-1.386(7)   \\
1.960370  && 0.1 && 	5.0  && 	-1.090214  && 	-1.541(2)   \\
1.411070  && 0.3 && 	5.0  && 	-1.238449  && 	-1.773(5)   \\
0.809079  && 1.0 && 	5.0  && 	-1.671462  && 	-2.312(4)   \\
0.259773  && 3.0 && 	5.0  && 	-2.770084  && 	-3.504(4)   \\
-0.342214  && 10.0 && 	5.0  && 	-6.376746  && 	-7.172(4)   \\
\hline\hline
\end{tabular}
\caption{Data of the equation of state presented in Fig.~\ref{fig:EOS}.}
\label{table:EOS_data}
\end{table}

\end{document}